\documentclass[conference]{IEEEtran}
\IEEEoverridecommandlockouts
\usepackage{color,soul}
\usepackage{enumitem}
\usepackage[dvipsnames]{xcolor}
\usepackage{graphicx}
\usepackage{multirow}
\usepackage{colortbl} 
\usepackage{subcaption}

 \usepackage{cite}
\usepackage{amsmath,amssymb,amsfonts}
\usepackage{graphicx}
\usepackage{textcomp}
\usepackage{subcaption}
 \usepackage{booktabs}
 \usepackage{multirow} 
\usepackage{soul}
\usepackage{tikz}
\usepackage{algorithm}
\usepackage{algpseudocode}

\usepackage{xcolor}
\usepackage{cuted}
\newtheorem{definition}{\textbf{Definition}}

\begin{document}
\title{Privacy-aware IoT Fall Detection Services \\For Aging in Place
}

\author{\IEEEauthorblockN{Abdallah Lakhdari\IEEEauthorrefmark{1},
Jiajie Li\IEEEauthorrefmark{2}, 
Amani Abusafia\IEEEauthorrefmark{3} and
Athman Bouguettaya\IEEEauthorrefmark{1}}
\IEEEauthorblockA{\IEEEauthorrefmark{1}School of Computer Science,
The University of Sydney, Australia}
\IEEEauthorblockA{\IEEEauthorrefmark{2}Department of Computer Science,
ETH Zürich, Switzerland}
\IEEEauthorblockA{\IEEEauthorrefmark{3}School of Computing and Information Systems,
The University of Melbourne, Australia\\
abdallah.lakhdari@sydney.edu.au, lijiaj@student.ethz.ch, amani.abusafia@unimelb.edu.au, athman.bouguettaya@sydney.edu.au}}



\maketitle

\begin{abstract}
Fall detection is critical to support the growing elderly population, projected to reach 2.1 billion by 2050. However, existing methods often face data scarcity challenges or compromise privacy. We propose a novel IoT-based Fall Detection as a Service (FDaaS) framework to assist the elderly in living independently and safely by accurately detecting falls. We design a service-oriented architecture that leverages Ultra-wideband (UWB) radar sensors as an IoT health-sensing service, ensuring privacy and minimal intrusion. We address the challenges of data scarcity by utilizing a Fall Detection Generative Pre-trained Transformer (FD-GPT) that uses augmentation techniques.  We developed a protocol to collect a comprehensive dataset of the elderly daily activities and fall events. This resulted in a real dataset that carefully mimics the elderly's routine. We rigorously evaluate and compare various models using this dataset. Experimental results show our approach achieves 90.72\% accuracy and 89.33\% precision in distinguishing between fall events and regular activities of daily living.


\end{abstract}


\section{Introduction}

The Internet of Things (IoT) enables everyday physical objects, or "things," to be connected to the Internet \cite{al2015internet}. These objects are often equipped with pervasive intelligence capabilities. IoT devices' capabilities may be abstracted as \textit{IoT services} \cite{chaki2021dynamic}. An IoT service has a set of functional and non-functional, i.e., quality of service (QoS) properties. For instance, a light bulb in a smart home is considered a light service, where its functional property is illumination, and its non-functional properties include luminous intensity, color, and connectivity. IoT services enable advanced applications such as smart cities, grids, and healthcare systems \cite{chaki2021dynamic}.

Smart homes leverage IoT services to enhance residents' \textit{convenience} and \textit{efficiency}\cite{chaki2021dynamic}. Recently, smart homes have been proposed as a solution to enable in-home quality aging, i.e., \textit{smart aging}\cite{Pal2018} \cite{pal2017smart}. Smart aging focuses on the composition of IoT services to monitor health, detect adverse events, and promote safety and independence for elderly individuals \cite{choi2019emerging}. A key component of smart aging is providing adaptive services that help the elderly maintain their independence. This includes fall detection systems, which are crucial for enhancing safety and enabling timely emergency responses.

Among these adaptive services, fall detection systems are particularly important due to the high risk and serious consequences of falls in elderly populations. The global population aged 60 and older is expected to reach 2.1 billion by 2050, leading to a significant increase in healthcare and caregiving costs \cite{organization_2022}. This growing elderly population makes efficient and adaptable fall detection solutions essential for reducing the burden on caregivers and improving safety in smart homes.

Existing fall detection systems primarily rely on IoT health-sensing services using wearable and vision-based devices. Wearable-based IoT services use accelerometers or gyroscopes to measure changes in velocity, acceleration, or orientation to determine abrupt changes, such as a fall event \cite{Ramachandran2020}. These systems have shown high accuracy, but they depend on the elderly consistently wearing devices, which is often unrealistic. Many elderly individuals resist or forget to wear these devices, leading to inconsistent performance. Vision-based services employ cameras and computer vision solutions to detect falls. However, these methods face adoption challenges due to privacy concerns \cite{Alkhatib2021}.\looseness=-1

\begin{figure*}[!t]
    \centering
    \includegraphics[width=0.8\linewidth]{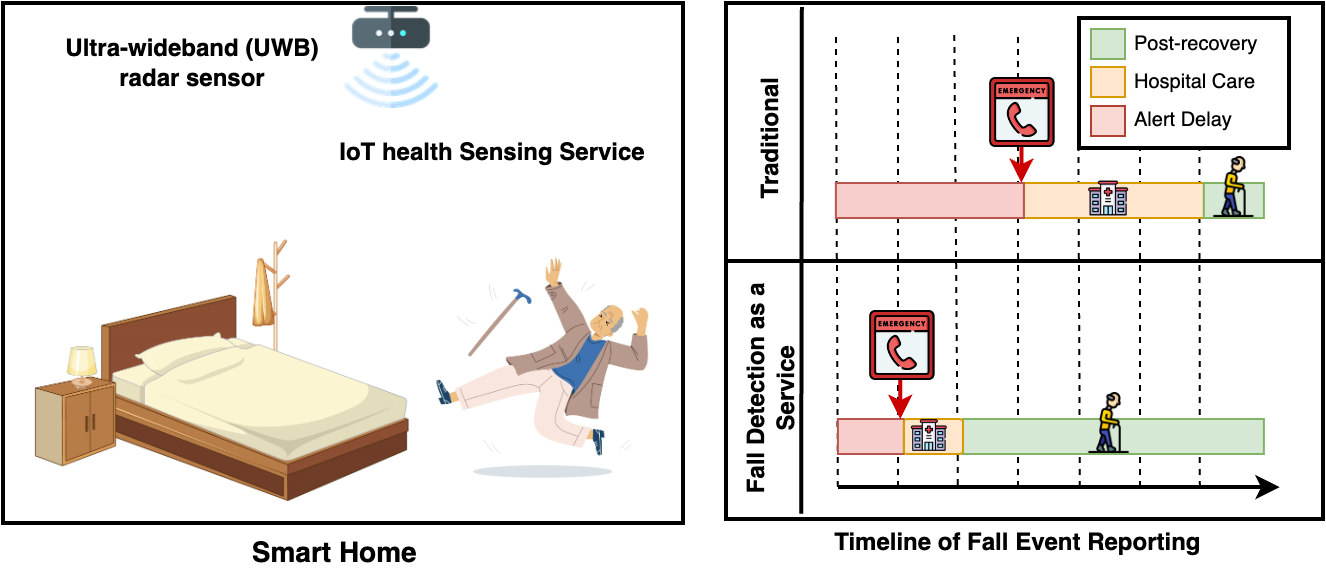}
    \caption{Example of the impact of fall detection services using unobtrusive IoT health sensing services}
    \label{fig:introImg}
\end{figure*}

Researchers have explored \textit{passive and unobtrusive ambient sensors} as a potential solution to these challenges. These systems minimize interference with daily routines and do not require the elderly to adapt to new technologies. Many existing unobtrusive frameworks, however, mainly rely on small self-collected datasets \cite{personalized_fall_detection}. This limits their scalability and effectiveness in real-world deployments\cite{CNN_video_based_fall_detection_high_accuracy, wearable_based_fall_detection_with_CNN, ambient_based_fall_detection_Doppler_CNN}. Additionally, most existing frameworks are not context-aware. They do not adapt their service delivery based on the resident’s needs or the changing environment. These systems apply the same detection model to all residents, regardless of their mobility, health conditions, or daily routines. For example, a high-risk resident may need faster response times. An active resident might benefit more from precise activity differentiation. These limitations reduce the reliability of fall detection systems. These limitations hinder the development of reliable fall detection systems. Therefore, a dynamic and scalable approach is needed to adapt services to the specific needs of residents and ensure consistent performance.

We propose Fall Detection-as-a-Service (FDaaS) as a dynamic and adaptive service for detecting falls in smart homes. The key functional aspect of FDaaS is determining whether a fall has occurred. The non-functional attributes (QoS), such as accuracy, precision, privacy, data efficiency, and service reliability, define the quality criteria for how the service operates within smart home environments. The service approach offers a significant alternative by leveraging service computing principles to reformulate fall detection in IoT-based smart homes as an optimization problem. This involves balancing non-functional attributes while considering the specific requirements of smart home elderly residents.

We propose a novel service-based, non-intrusive fall detection framework to deliver FDaaS effectively (See Fig.\ref{fig:introImg}). The framework integrates IoT health-sensing services using Ultra-wideband (UWB) radar sensors for in-home elderly monitoring. The UWB radar provides a non-intrusive method for capturing vital signs while maintaining privacy. The framework employs a Fall Detection Generative Pre-trained Transformer (FD-GPT) to address data scarcity and imbalance through augmentation. Generative models enable FD-GPT to learn and adapt to specific user behaviors, enhancing performance over traditional systems \cite{bansal2024transforming}. Additionally, the framework incorporates FD-prompt, a rule-based approach for context-aware service delivery. This approach adapts the service to the specific needs of elderly residents, considering factors such as age group, health condition, and available resources. For example, FDaaS can adjust model selection criteria for elderly residents with critical health conditions requiring tailored monitoring. This context-aware framework ensures FDaaS integrates seamlessly with other smart home services, such as emergency response systems and health monitoring platforms. The main contributions of this paper are:

\begin{enumerate}
  \item   A novel Fall Detection-as-a-Service (FDaaS) service model.

\item A service-based framework for context-aware fall detection.

\item An optimization strategy using generative AI techniques to enhance the reliability of fall detection models.
\end{enumerate}

\subsubsection*{\textbf{Motivation Scenario}} 

Assume an elderly resident living in a smart home equipped with various IoT services to support in-home quality aging (See Fig.\ref{fig:introImg}). Among these services is an IoT health sensing service that monitors the resident's health data using an unobtrusive Ultra-wideband (UWB) radar sensor. This radar sensor continuously gathers information on the resident's movements and activities. The collected data is then pushed to the edge for real-time analysis.

A Fall Detection-as-a-Service (FDaaS) framework operates at the edge to accurately detect falls. The framework integrates a Fall Detection Generative Pre-trained Transformer (FD-GPT) to address data imbalance through augmentation techniques. FD-GPT uses generative models to create synthetic data, improving the model's reliability even with limited training datasets. Additionally, the framework ensures privacy by using non-intrusive sensing methods and processing data locally, minimizing the exposure of sensitive health information.

The impact of FDaaS is illustrated in Fig.\ref{fig:introImg}, demonstrating how accurate fall detection as a service can reduce emergency response times. Traditional methods often lead to longer alert delays, resulting in extended hospital care and slower recovery. In contrast, FDaaS enables immediate fall detection and automatic alerting to emergency services. This faster response minimizes hospital stays and promotes quicker recovery, helping the resident maintain independence.


FDaaS employs a context-aware approach to optimize fall detection services based on the specific needs of elderly residents. The system uses a rule-based FD-prompt to dynamically adjust service parameters according to the resident's age, health condition, and available resources. For instance, a high-risk resident with limited mobility may require the service to prioritize accuracy and reliability, ensuring that even minor falls are detected promptly. Conversely, for a more active resident, FDaaS may adjust its model to reduce false positives, allowing greater freedom of movement without unnecessary alerts. By continuously adapting to the resident’s context, FDaaS enhances not only the safety and reliability of fall detection but also supports a higher quality of life by providing a personalized and unobtrusive service experience.

\section{Preliminaries and Problem Formulation}
This section introduces the notions of IoT health sensing service (HS) and IoT sensing service events (HSE). These definitions are adapted from \cite{chaki2021dynamic}. We also introduce the definition of Fall Detection-as-a-Service (FDaaS) which aims to detect falls from IoT sensing service events.

\begin{definition} 
\label{HS_def}
\textbf{IoT Health Sensing Service (HS)} is a tuple $<id, Rid, F, Q>$, where:

\begin{itemize}
    \item \textbf{id} is the service ID.
    \item \textbf{Rid} is the resident ID.
    \item \textbf{F} is the function of sensing and collecting health data using the radar.
    \item \textbf{Q} is a set of non-functional attributes, defined as $<L, T, D>$, where:
    \begin{itemize}[noitemsep,nosep,leftmargin=1pt,labelsep=1pt,itemindent=0pt, labelwidth=*] 
        \item $L$ is the sensor's location (e.g., the bedroom).
        \item $T$ is the time duration of the service, defined with a start and end time [st, et].
           \item $D$ is a continuous data stream collected by the radar sensor. In our context, the radar sensor collects four values every second as $<hr, br, d, ps>$, where:
        \begin{itemize}
            \item $hr$ is the resident's heart rate, e.g., 54 times/min (heartbeats per minute).
            \item $br$ is the resident's breathing rate, e.g., 8 Times/M (breaths per minute).
            \item $d$ is the distance between the radar and the detected subject, e.g., 1.72 m.
            \item $ps$ is the physiological state, which is a categorical variable with the following values: 0 (no presence), 1 (constant calm), 2 (intermittent), 3 (minor movement), and 4 (continued).
           
\end{itemize}
    \end{itemize}
\end{itemize}

\end{definition}

\begin{definition}\label{HSE_def}\textbf{IoT Health Sensing Service Event (HSE)} represents a discrete snapshot of the health data captured at a particular time and location by the IoT Health Sensing Service (HS). An HSE captures the observed health data, the exact time, and the location during the service operation.

An HSE is defined as a tuple $<HSE_{id}, data, TS, L>$, where:

\begin{itemize}
    \item \textbf{$HSE_{id}$} is the unique identifier of the health sensing event.
    \item \textbf{$data$}  is a snapshot of the health data as defined in Definition \ref{HS_def}. It includes heart rate, breathing rate, distance to the subject, and physiological state.
    \item \textbf{$TS$} is a sequence of timestamps \( [t_1, t_2, ..., t_n] \) indicating when the data was captured.
    \item \textbf{$L$} is the sensor’s location (e.g., bedroom).
\end{itemize}

An \textbf{IoT Health Sensing Service Event Sequence (HSES)} is an ordered collection of health sensing events 
\( \{ HSE_1, HSE_2, \ldots, HSE_k \} \) that represent the chronological record of observations.
\end{definition}

\begin{definition}\label{FDAAS_def}
\textbf{Fall Detection-as-a-Service (FDaaS)} is a tuple $<id, F, Q>$, where:

\begin{itemize}
    \item \textbf{id} is the service ID.
    \item \textbf{F} is the function of ascertaining whether a fall has occurred.
    \item \textbf{Q} is a set of non-functional attributes, defined as $<A, F1, T, P, R>$, where:
    \begin{itemize}
        \item $A$ is the required accuracy level (e.g., 95\%).
        \item $F1$ is the F1 score, balancing precision and recall, particularly useful for imbalanced datasets.
        \item $T$ is the acceptable latency for fall detection (e.g., max 2 seconds).
        \item $P$ is the privacy level (e.g., data anonymization).
        \item $R$ is the reliability of the service (e.g., 99.9\% uptime).
    \end{itemize}
\end{itemize}
\end{definition}

 In this paper, we focus primarily on the \textbf{accuracy} ($A$) of fall detection as the key Quality of Service (QoS) attribute for service selection within the FDaaS framework.

\begin{figure*}[!t]
    \centering
    \includegraphics[width=0.8\linewidth]{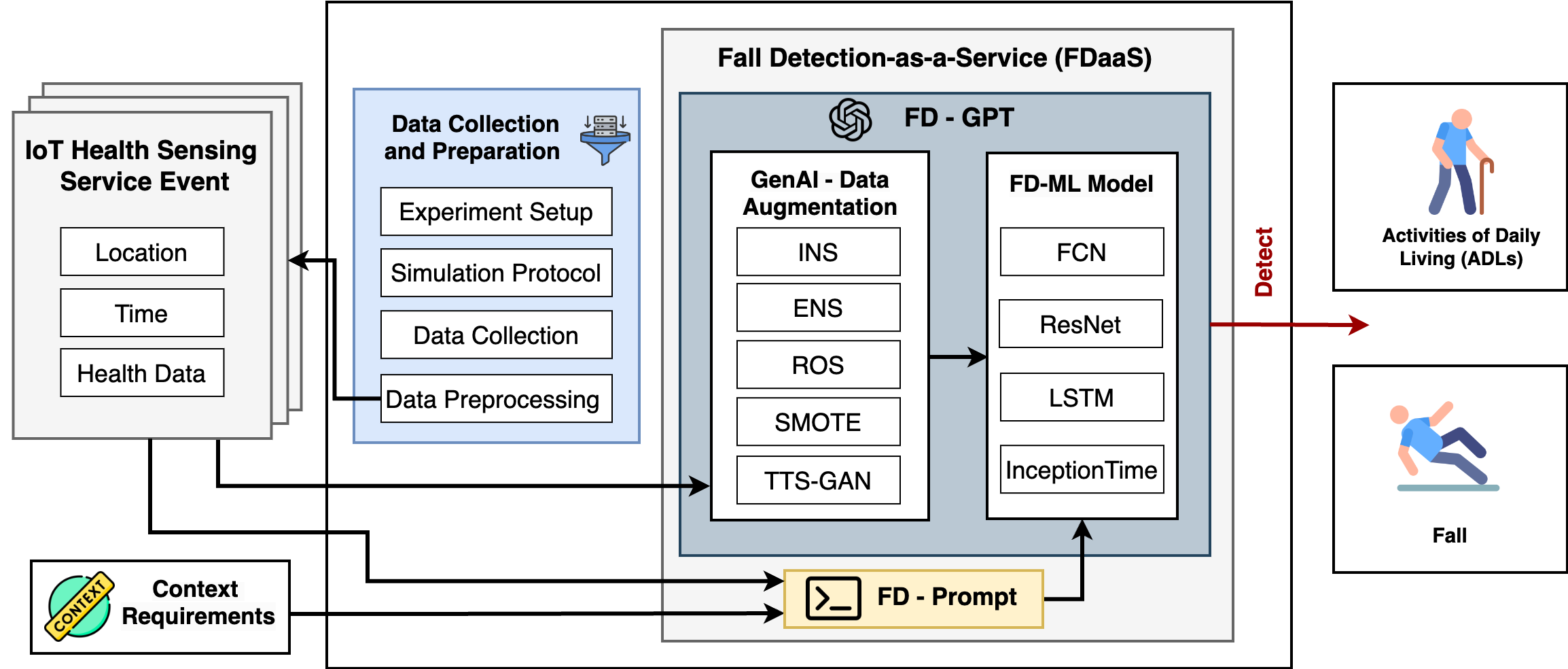}
    \caption{An overview of the fall detection service system}
    \label{fig:FDSOverview}
\end{figure*}

\subsection{Problem Statement}
Given an IoT Health Sensing Service (\textit{HS})  that monitors the health of elderly residents using a non-intrusive UWB radar sensor. This service collects health data such as heart rate, breathing rate, distance, and physiological state. An IoT Health Sensing Service Event (\textit{HSE}) captures this data, and IoT Health Sensing Service Event Sequences (\textit{HSES}) record the history of these events. The goal of this paper is to design and develop FDaaS. The \textit{functional part} of FDaaS \( F(HSES) \) accurately \textit{detects falls} by (1) leveraging IoT health sensing service event sequences and (2) considering the context requirements as \textit{non-functional requirements}, i.e., The function should address challenges, including data imbalance, privacy concerns, and real-time processing.


\section{Fall Detection as a Service Framework}\label{framework}
This section presents the fall detection system using IoT health sensing services (see Fig.\ref{fig:FDSOverview}). Our system aims to classify activities of daily living (ADL) and fall events. The system consists of two main phases (1) Data Collection and Preparation, and (2) the Fall Detection as a Service (FDaaS) Framework. A recent comprehensive fall dataset was published to address limitations in existing publicly available wearable-based fall datasets \cite{FallAllD_dataset}. However, their protocol did not consider the behavioral characteristics of older adults or the common types of falls that occur in real-world settings. Therefore,  we conducted experiments to collect a new, real-labeled dataset of falls and actions specifically for elderly individuals. As the dataset collection is not the main focus of this paper, further details are discussed in the experiments section (See Sec. \ref{experiments}). However, this section provides an overview of the recorded data and its role in the fall detection framework.

The FDaaS framework consists of two main components: (1) the Fall Detection Generative Pre-trained Transformer (FD-GPT) and (2) the Fall Detection (FD) Prompt. The FD-GPT consists of two phases: data augmentation and Machine Learning (FD-ML) model training. The data augmentation phase aims to balance the dataset, enabling the fall detection models to learn discriminative patterns from the majority class (i.e., ADLs) and the minority class (i.e., falls). The FD-ML model training phase employs several proposed models to differentiate between falls and ADL events. The FD-GPT is trained offline and periodically updated with new data. In real-time, the edge-based FD Prompt invokes the FD-GPT to recognize falls and ADLs. The prompts consider the environment's context requirement when invoking FD-GPT, such as accuracy, latency, etc. For instance, if the elderly resident is over 90 years old, then latency may be more critical than accuracy. Each component of the framework is detailed below.\looseness=-1

\subsection{Data Overview}

\begin{figure}[!t]
\centering
    \includegraphics[width=0.6\linewidth]{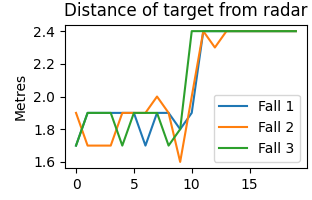}
    \caption{Example of the collected time series recordings of the distance for one subject who tripped and fell three times }
    \label{fig:fallplots}
\end{figure}

The used Ultra-wideband (UWB) radar sensor in our context records four attributes as the QoS attributes of the  IoT Health Sensing Service ($HS$) (See Def. \ref{HS_def}). These attributes are distance, heart rate, breathing rate, and physiological movement. Figure \ref{fig:fallplots} shows an example of the recordings for the distance attribute for one subject who tripped and fell three times. The IoT Health Sensing Service Event Sequence (\textit{HSES}) represents a stream of data recordings of the elderly daily activities and falls in the smart house (See Table \ref{tab:simulation_protocol} for examples of activities and falls).\looseness=-1

\begin{table}[!t]
\centering
 \captionof{table}{ADLs and Fall Types}

\resizebox{0.9\linewidth}{!}{%
        \renewcommand{\arraystretch}{1}
\begin{tabular}{|c|ll}
\hline
\rowcolor[HTML]{9B9B9B} 
\textbf{Category}                                               & \multicolumn{1}{c|}{\cellcolor[HTML]{9B9B9B}\textbf{Short Description}} & \multicolumn{1}{c|}{\cellcolor[HTML]{9B9B9B}\textbf{Code}} \\ \hline
\cellcolor[HTML]{C0C0C0}                                        & \multicolumn{1}{l|}{Standing}                                           & \multicolumn{1}{l|}{STA}                                   \\ \cline{2-3} 
\cellcolor[HTML]{C0C0C0}                                        & \multicolumn{1}{l|}{Sitting on a bed edge}                              & \multicolumn{1}{l|}{SOB}                                   \\ \cline{2-3} 
\cellcolor[HTML]{C0C0C0}                                        & \multicolumn{1}{l|}{Sitting on a chair}                                 & \multicolumn{1}{l|}{SOC}                                   \\ \cline{2-3} 
\cellcolor[HTML]{C0C0C0}                                        & \multicolumn{1}{l|}{Lying on a bed}                                     & \multicolumn{1}{l|}{LOB}                                   \\ \cline{2-3} 
\cellcolor[HTML]{C0C0C0}                                          & \multicolumn{1}{l|}{Lying on the ground}                                & \multicolumn{1}{l|}{LOG}                                   \\ \cline{2-3} 
\cellcolor[HTML]{C0C0C0}                                        & \multicolumn{1}{l|}{Walking}                                            & \multicolumn{1}{l|}{WLK}                                   \\ \cline{2-3} 
\cellcolor[HTML]{C0C0C0}                                        & \multicolumn{1}{l|}{Sitting down on a chair}                            & \multicolumn{1}{l|}{SDC}                                   \\ \cline{2-3} 
\cellcolor[HTML]{C0C0C0}                                        & \multicolumn{1}{l|}{Standing up from a chair}                           & \multicolumn{1}{l|}{SUC}                                   \\ \cline{2-3} 
\cellcolor[HTML]{C0C0C0}                                        & \multicolumn{1}{l|}{Sitting down on a bed edge}                         & \multicolumn{1}{l|}{SEB}                                   \\ \cline{2-3} 
\cellcolor[HTML]{C0C0C0}                                        & \multicolumn{1}{l|}{Putting legs onto a bed}                            & \multicolumn{1}{l|}{TOB}                                   \\ \cline{2-3} 
\cellcolor[HTML]{C0C0C0}                                        & \multicolumn{1}{l|}{Lying down on a bed}                                & \multicolumn{1}{l|}{LDB}                                   \\ \cline{2-3} 
\cellcolor[HTML]{C0C0C0}                                        & \multicolumn{1}{l|}{Rising up from a bed}                               & \multicolumn{1}{l|}{RIB}                                   \\ \cline{2-3} 
\cellcolor[HTML]{C0C0C0}                                        & \multicolumn{1}{l|}{Putting legs out of a bed}                          & \multicolumn{1}{l|}{TFR}                                   \\ \cline{2-3} 
\cellcolor[HTML]{C0C0C0}                                        & \multicolumn{1}{l|}{Standing up from a bed edge}                        & \multicolumn{1}{l|}{RIB}                                   \\ \cline{2-3} 
\multirow{-15}{*}{\cellcolor[HTML]{C0C0C0}\textbf{ADLs}}      & \multicolumn{1}{l|}{Standing up from the ground}                        & \multicolumn{1}{l|}{SUG}                                   \\ \hline
\cellcolor[HTML]{C0C0C0}                                        & \multicolumn{1}{l|}{Falling out of a bed}                               & \multicolumn{1}{l|}{FOB}                                   \\ \cline{2-3} 
\cellcolor[HTML]{C0C0C0}                                        & \multicolumn{1}{l|}{Falling by tripping}                                & \multicolumn{1}{l|}{FTR}                                   \\ \cline{2-3} 
\cellcolor[HTML]{C0C0C0}                                        & \multicolumn{1}{l|}{Falling while sitting down}                         & \multicolumn{1}{l|}{FST}                                   \\ \cline{2-3} 
\cellcolor[HTML]{C0C0C0}                                        & \multicolumn{1}{l|}{Falling while standing up}                          & \multicolumn{1}{l|}{FSU}                                   \\ \cline{2-3} 
\multirow{-5}{*}{\cellcolor[HTML]{C0C0C0}\textbf{Falls}}      & \multicolumn{1}{l|}{Falling while turning}
                    & \multicolumn{1}{l|}{FTU}                                 \\ \hline
\end{tabular}
}
\label{tab:simulation_protocol}
\end{table}


\subsection{Fall Detection Generative Pre-trained Transformer (FD-GPT)}

The Fall Detection Generative Pre-trained Transformer (FD-GPT) is a critical component of our fall detection framework. It leverages advanced machine learning and generative AI techniques to improve the accuracy and robustness of fall detection models. FD-GPT integrates data augmentation methods and deep learning algorithms to process and analyze health data captured by IoT Health Sensing Services. By using generative models, e.g., GAN (see Fig. \ref{fig:fall_data_augmentation} ), FD-GPT can synthesize realistic fall scenarios and enhance the training datasets. This approach addresses the class imbalance issue commonly encountered in fall detection frameworks. Consequently, the system can accurately distinguish between Activities of Daily Living (ADLs) and various types of falls, improving overall performance and reliability.

\subsubsection{Data Augmentation}\label{data_aug}


Falls are rare events in daily life, which results in imbalanced sample sizes between falls and Activities of Daily Living (ADL) events in collected datasets \cite{personalized_fall_detection}. As a result, existing fall detection systems often fail when customizing their models with personalized datasets. This failure occurs because the models tend to classify all samples as ADL,i.e., the majority class. The imbalance issue can be addressed using data augmentation techniques \cite{he2008adasyn}\cite{islam2022knnor}.

\begin{figure}[!t]
    \centering
     \includegraphics[width=\linewidth]{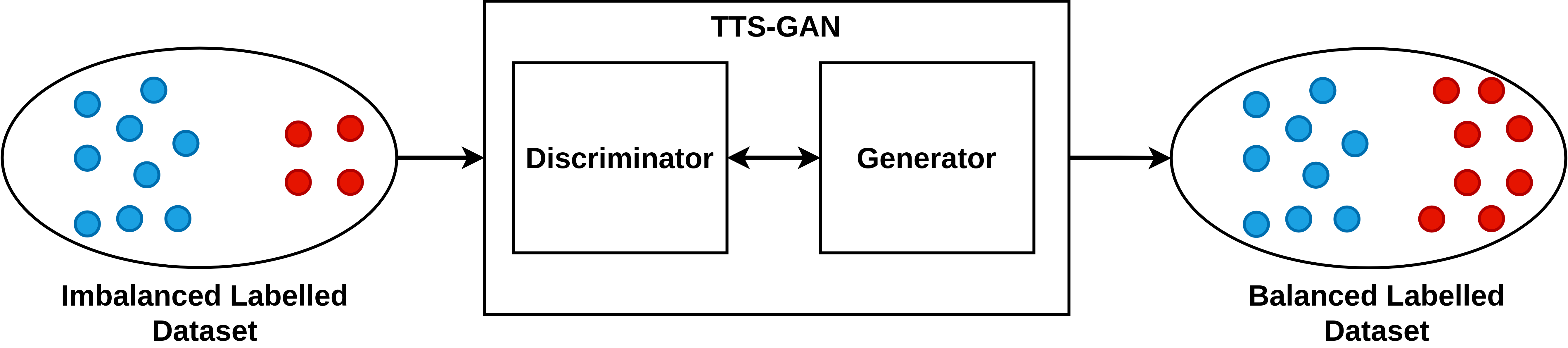}
    \caption{TTS-GAN-based data augmentation}
    \label{fig:fall_data_augmentation}
\end{figure}

We experimented with multiple augmentation methods, including weighted loss strategies (INS, ENS), random oversampling (ROS), and SMOTE, to address this issue \cite{he2008adasyn}\cite{ENS_weighted_loss}. Additionally, we used a state-of-the-art GAN model for time series data augmentation, namely TTS-GAN \cite{TTS_GAN}, to generate artificial fall samples. The goal is to approximate the distribution of fall data and generate synthetic samples to enhance model learning. These techniques help the fall detector learn fall patterns more effectively and improve performance, as illustrated in Fig. \ref{fig:fall_data_augmentation}.\looseness=-1

\subsubsection{Fall detection Machine learning Model}
\label{FDML}

This step focuses on building a machine-learning model for fall detection. The model will identify whether a fall has happened by distinguishing if an event is an ADL or a fall. Most fall detectors in the literature today employ deep neural networks that can automatically learn features from data \cite{FCN_fall_detection}\cite{ResNet_fall_detection}. Therefore, we have applied four models to detect falls: FCN, ResNet, LSTM, and InceptionTime. InceptionTime is a state-of-the-art time series classification model that has produced remarkable results on public time series classification datasets \cite{InceptionTime_paper}. However, it has not yet been employed in fall detection systems.\looseness=-1

\subsection{Fall Detection Prompt}

The fall detection prompt considers the resident’s context requirements when \textit{invoking} FD-GPT, such as accuracy, latency, and other relevant factors. These requirements are essential for tailoring the system to individual needs. For example, if a resident is over 90 years old with a critical health condition, the system prioritizes latency over accuracy to ensure an immediate response. In contrast, for a younger resident with a stable condition, the system prioritizes accuracy to reduce false positives.\looseness=-1

The system improves performance by adjusting fall detection parameters based on contextual factors. The prompts help FD-GPT make real-time, context-aware decisions, balancing multiple requirements for personalized fall detection. FD-Prompt employs a rule-based model selection approach, guided by health experts, to dynamically choose the appropriate fall detection model. Each selection follows predefined rules that account for age, health condition, and available resources. While we propose FD-Prompt as a component of the FDaaS framework, designing its full rule-based optimization approach will be addressed in future work. An example of model selection criteria is shown in Algorithm \ref{alg:model_selection}. These criteria are proposed based on logical assumptions rather than expert evaluation or experimental validation.

\begin{algorithm}[!t]
\caption{Model Selection for Fall Detection Service}
\label{alg:model_selection}
\begin{algorithmic}[1]
\Require age\_group, health\_condition, resource\_availability
\Ensure selected\_model

\If {health\_condition = ‘‘Critical’’}
    \If {resource\_availability = ‘‘Limited’’}
        \State selected\_model = ‘‘FCN’’
    \Else
        \State selected\_model = ‘‘ResNet’’
    \EndIf
\ElsIf {age\_group = ‘‘Elderly (80+ years)’’}
    \If {resource\_availability = ‘‘Limited’’}
        \State selected\_model = ‘‘LSTM’’
    \Else
        \State selected\_model = ‘‘InceptionTime’’
    \EndIf
\Else
    \If {resource\_availability = ‘‘Limited’’}
        \State selected\_model = ‘‘LSTM’’
    \Else
        \State selected\_model = ‘‘InceptionTime’’
    \EndIf
\EndIf

\State \Return selected\_model
\end{algorithmic}
\end{algorithm}

\section{Experimental Results and Discussion}\label{experiments}

We conducted experiments to evaluate both components of the framework: data augmentation and fall detection. We compare our TTS-GAN augmentation method with four additional techniques: (1) Inverse of Number of Samples (INS): INS is a simple weighted loss function that mitigates the impact of imbalanced datasets by weighting samples inversely to their class frequency. (2) Effective Number of Samples (ENS): Unlike INS, ENS considers the 'effective number of samples' to better represent critical samples for classification. This method weights samples inversely to the effective number in their class, enhancing performance on imbalanced datasets. (3) Random Oversampling (ROS): ROS addresses data imbalance by resampling more points from the minority class with replacement, equalizing class sizes. (4) Synthetic Minority Oversampling Technique (SMOTE): SMOTE generates synthetic samples close to real ones in high-dimensional space, effectively handling data imbalance. We use these augmentation methods with the fall detection models mentioned in Sec. \ref{FDML}, namely, FCN, ResNet, LSTM, and InceptionTime. In what follows, we present the dataset used,  introduce the hyperparameters settings, discuss the metrics applied, and share the findings of the experiments.

\subsection{Data Collection and Preparation}
As discussed in Sec. \ref{framework}, the details of the data collection and preparations will be presented in this section. The goal of this phase is to collect labeled sequences of falls and actions for the elderly. The data collection process consists of five components: (1) experiment setup, (2) simulation protocol, (3) data collection, and (4) preprocessing the data. In the following subsections, we present each component in detail.

\subsubsection{Experiments Setup}

In this subsection, we present the main components used to conduct our experiments: IR-UWB radar, the experimental environment, and the participants.

\begin{table*}[!t]
\centering
\caption{The description of four signals provided by our UWB radar}
\resizebox{\textwidth}{!}{%
\begin{tabular}{@{}cll@{}}
\toprule
\textbf{Signal}                               & \multicolumn{2}{c}{\textbf{Description}}                                                                                                                                                                                                                                                          \\ \midrule
\textbf{Heart Rate}                           & \multicolumn{2}{l}{The number of heartbeats of the detected subject per minute (e.g., 54 Times/M)}                                                                                                                                                                                                \\
\textbf{Breathing Rate}                       & \multicolumn{2}{l}{The number of breaths of the detected subject per minute (e.g, 8 Times/M)}                                                                                                                                                                                                     \\
\textbf{Distance}                             & \multicolumn{2}{l}{The distance between the radar and the detected subject (e.g., 1.72 m)}                                                                                                                                                                                                        \\
\multirow{2}{*}{\textbf{Physiological State}} & \multicolumn{2}{l}{\multirow{2}{*}{\begin{tabular}[c]{@{}l@{}}The intensity of the activity of the detected subject. This feature is a categorical variable and has 5 different values:\\ 0 (NO PRESENCE) , 1 (CONSTANT CALM), 2 (INTERMITTENT), 3 (MINOR MOVEMENT), 4 (CONTINUED)\end{tabular}}} \\
                                              & \multicolumn{2}{l}{}                                                                                                                                                                                                                                                                              \\ \bottomrule
\end{tabular}%
}

\label{tab:UWB_features}
\end{table*}

\subsubsection*{\textbf{1.1 IR-UWB Sensor}} We used an off-the-shelf UWB radar sensor to prioritize user privacy and deployment simplicity~\footnote{intelicare.com.au}. The radar sensor produces short radio frequency pulses. These pulses are emitted by the sensor's antenna to the object. Then reflected back to the receiving antenna, producing a receiving signal. The radar sensor can monitor four types of signals: heartbeat, breathing rate, distance, and physiological state. The physiological state encompasses the subject's activity intensity, categorized into five levels ranging from 0 (completely stationary) to 4 (exercising at high intensity). Figure \ref{fig:fallplots} shows an example of the recordings for the distance attribute for one subject who tripped and fell three times. A comprehensive overview of each signal is provided in Table \ref{tab:UWB_features}.

\begin{figure}[!t]
    \centering
    \includegraphics[width=\linewidth]{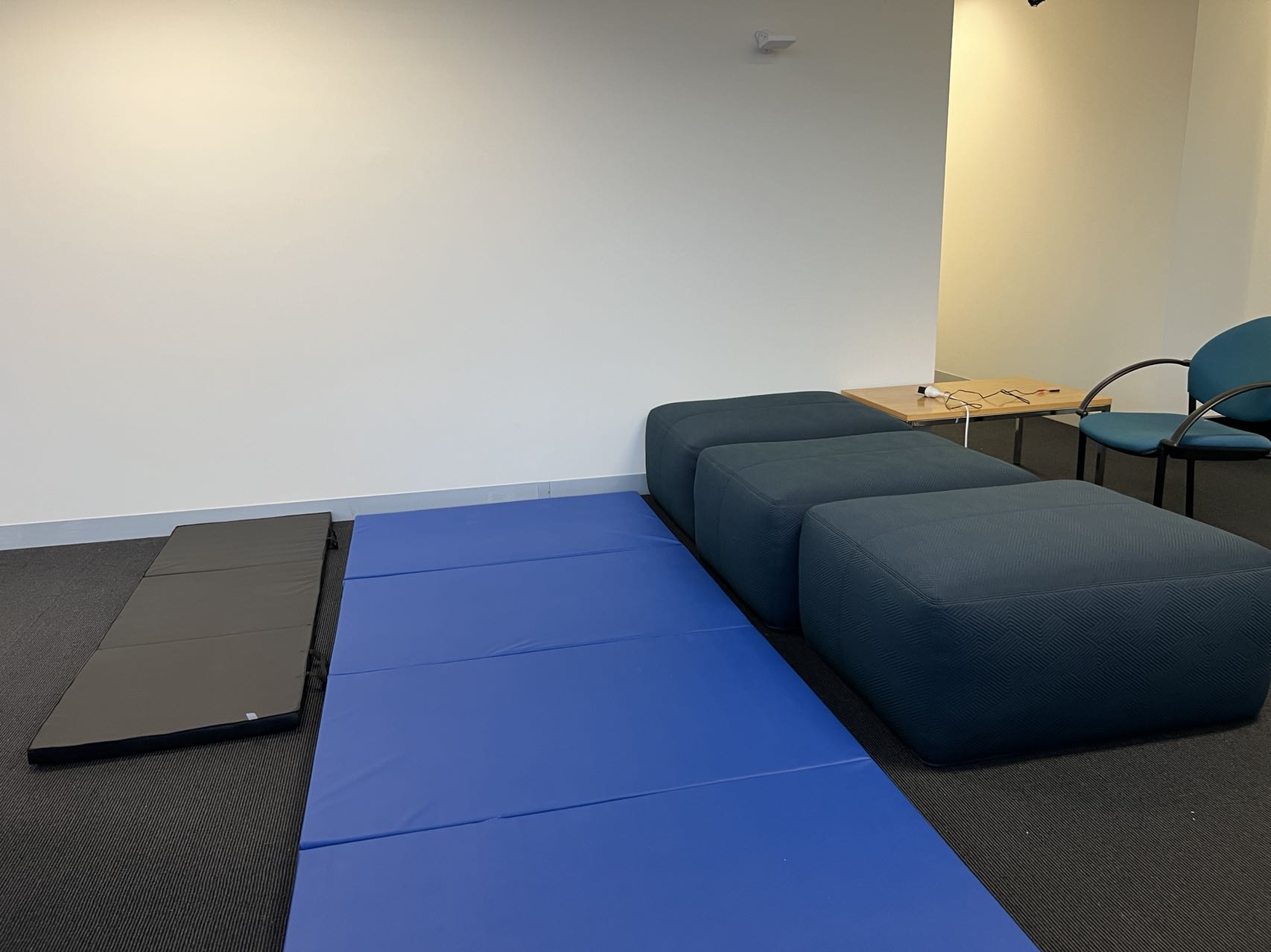}
    \caption{The bedroom experiment environment}
    \label{fig:images_of_bedroom_experiment_environment}
\end{figure}

\subsubsection*{\textbf{1.2 Bedroom Experiment Environment}} \label{subsec:experiment_environment}
The best application scenario for UWB radar is in a bedroom, so we built a simple bedroom environment to experiment with. Our bedroom has a bed, bedside table, chair, and protective pads on the ground, as shown in Fig. \ref{fig:images_of_bedroom_experiment_environment}. We followed the manual's recommendation to mount the UWB radar above the head of the bed for accurate measurements. The experiment was also conducted in a room near a more secluded area of the building to limit interference from passersby around the room. After the radar was deployed, we calibrated the radar using the software provided by the manufacturer. The detailed configuration of the UWB radar in the experiment is summarized in the Table \ref{tab:UWB_configuration}.\looseness=-1


\begin{table}[htp]
\centering
\resizebox{0.8\linewidth}{!}{%
\begin{tabular}{@{}cc@{}}
\toprule
\multicolumn{1}{l}{}               & \textbf{Configuration Value} \\ \midrule
\textbf{Minimum Sensing Frequency} & 1 m                          \\
\textbf{Maximum Sensing Frequency} & 3 m                          \\
\textbf{Sensitivity}               & 6                            \\
\textbf{Installation Height}       & 2 m                          \\ \bottomrule
\end{tabular}%
}
\caption{The configuration of the UWB radar}
\label{tab:UWB_configuration}
\end{table}

\subsubsection*{\textbf{1.4 Participants}} \label{subsec:participants}

Although the elderly are the ideal subjects, conducting fall experiments with them presents significant ethical and safety concerns. Therefore, we hired younger actors who were trained to accurately mimic the movements and behavioral patterns of elderly individuals. The actors have access to recorded examples of actions and falls to emulate. We ran fall experiments and collected data from 10 participants: 5 females and 5 males. Participants had an average age of 24.1 (24.1 \(\pm\) 3.7, range: 21-32), an average height of 1.73 m (1.73 \(\pm\) 0.06 m, range: 1.65-1.88 m), and an average mass of 76.3 kg (76.3 \(\pm\) 23.1 kg, range: 54-130 kg). Each participant performed all the activities presented in table \ref{tab:simulation_protocol} at least three times. After removing events influenced by noise or sensor failures, we have 250 fall samples and 8880 ADLs samples in the dataset.

\begin{figure*}[!t]
    \centering
   \includegraphics[width=0.8\linewidth]{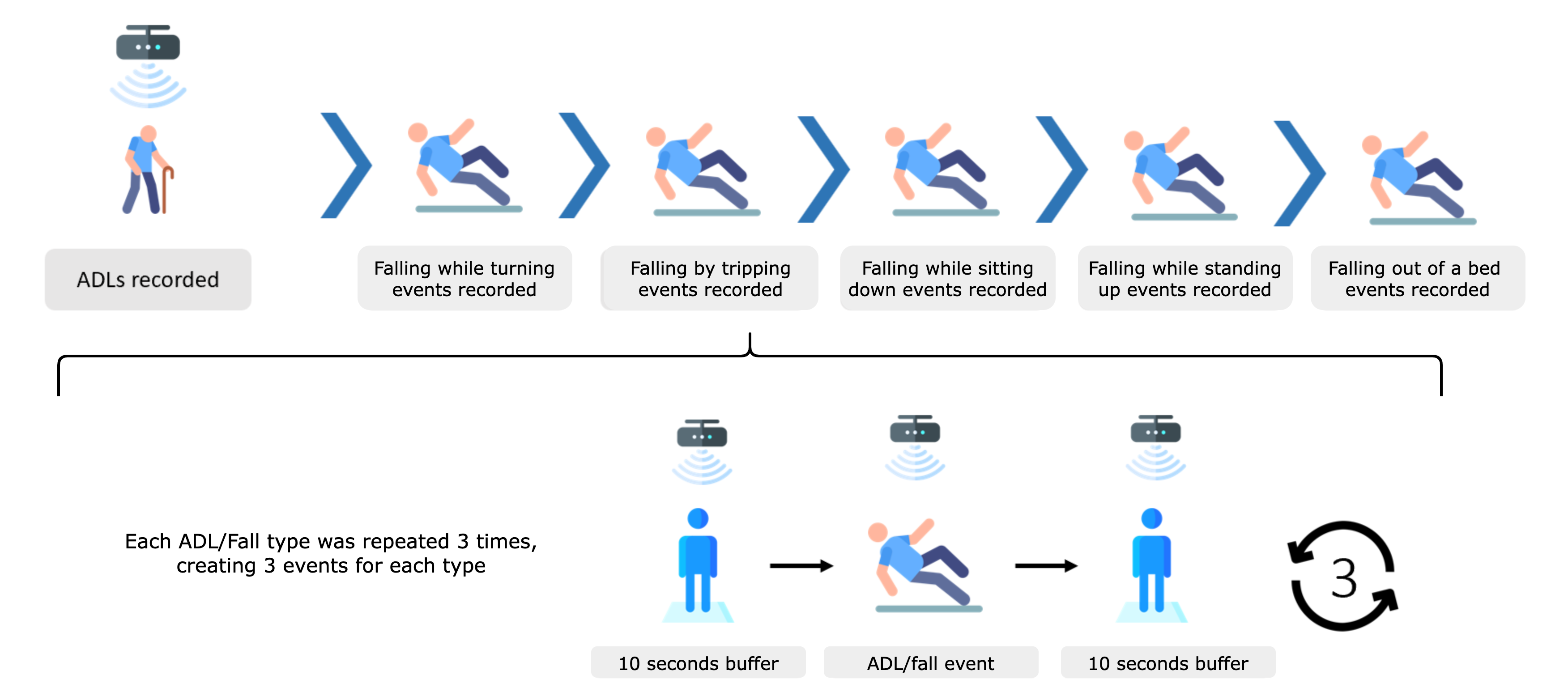}
    \caption{Experiment process: ADLS were recorded first, followed by each fall type. Each fall type was repeated three times to collect 3 fall events each.}
    \label{fig:process}
\end{figure*}

\begin{figure*}[!t]
    \centering

    \begin{subfigure}[b]{0.23\textwidth}
        \includegraphics[width=\textwidth]{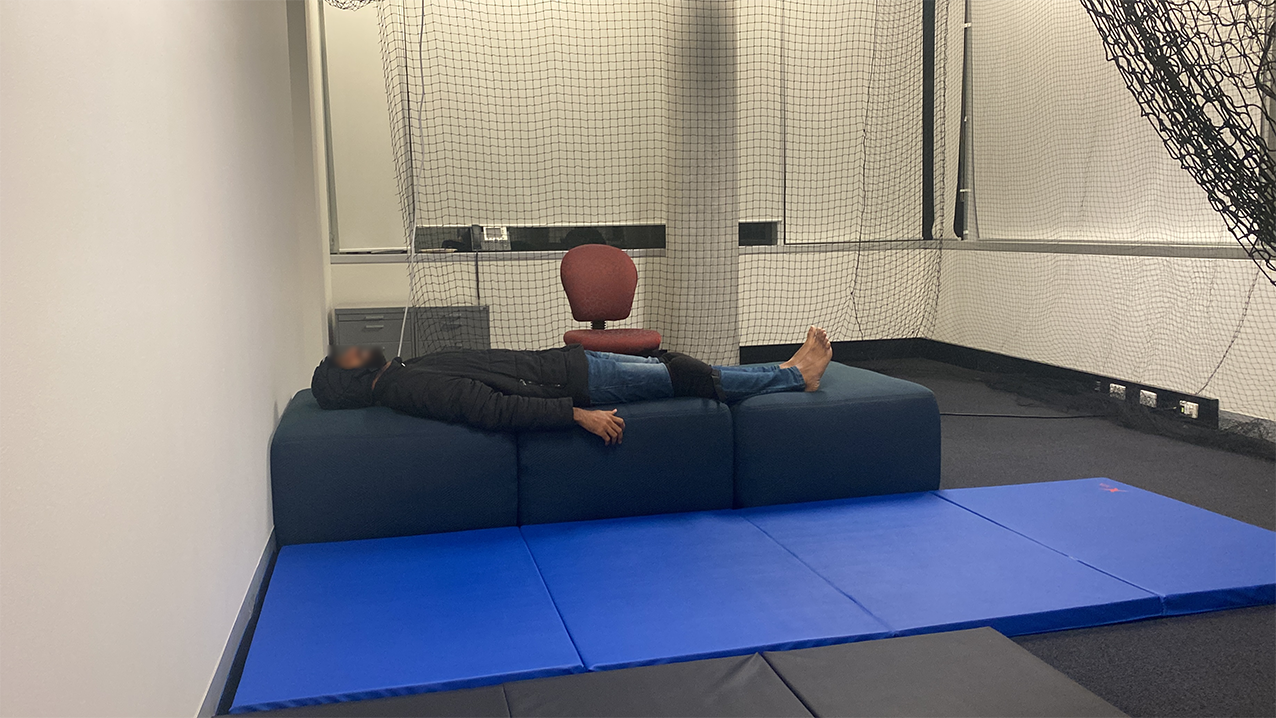}
        \caption{ADL: lying on a bed}
    \end{subfigure}
    \begin{subfigure}[b]{0.23\textwidth}
        \includegraphics[width=\textwidth]{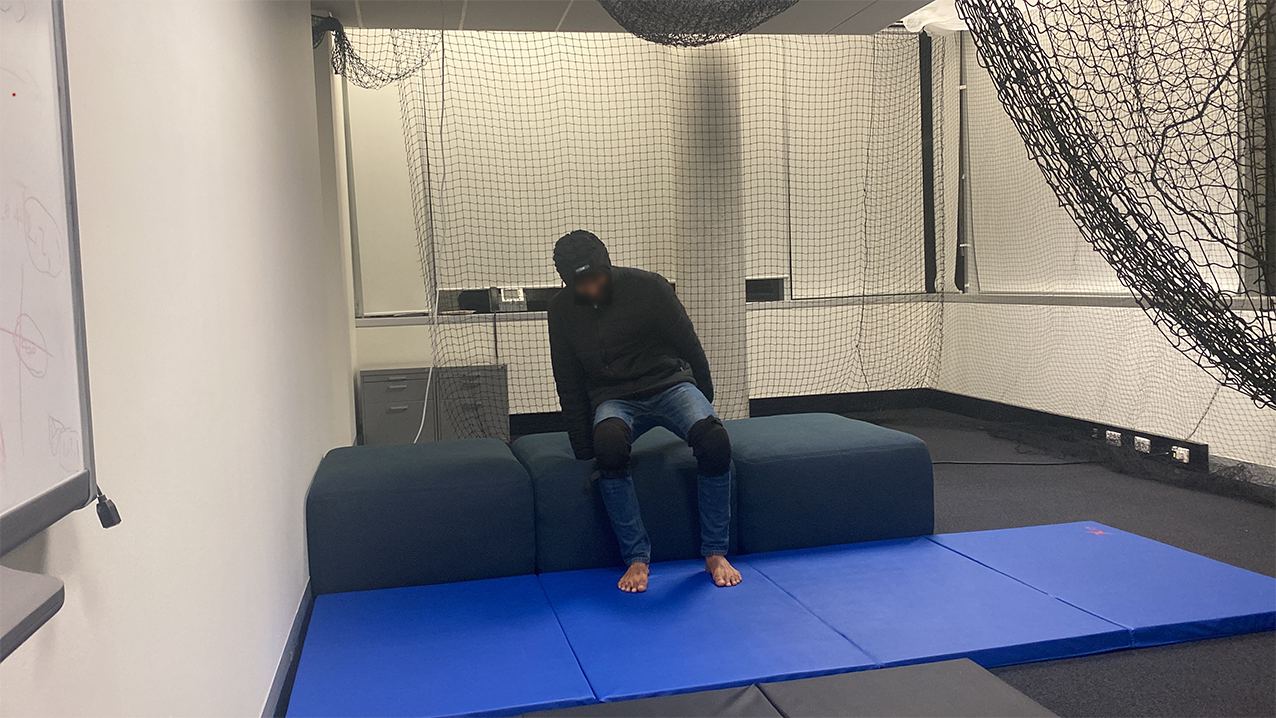}
        \caption{ADL: sitting down on a bed}
    \end{subfigure}
    \begin{subfigure}[b]{0.23\textwidth}
        \includegraphics[width=\textwidth]{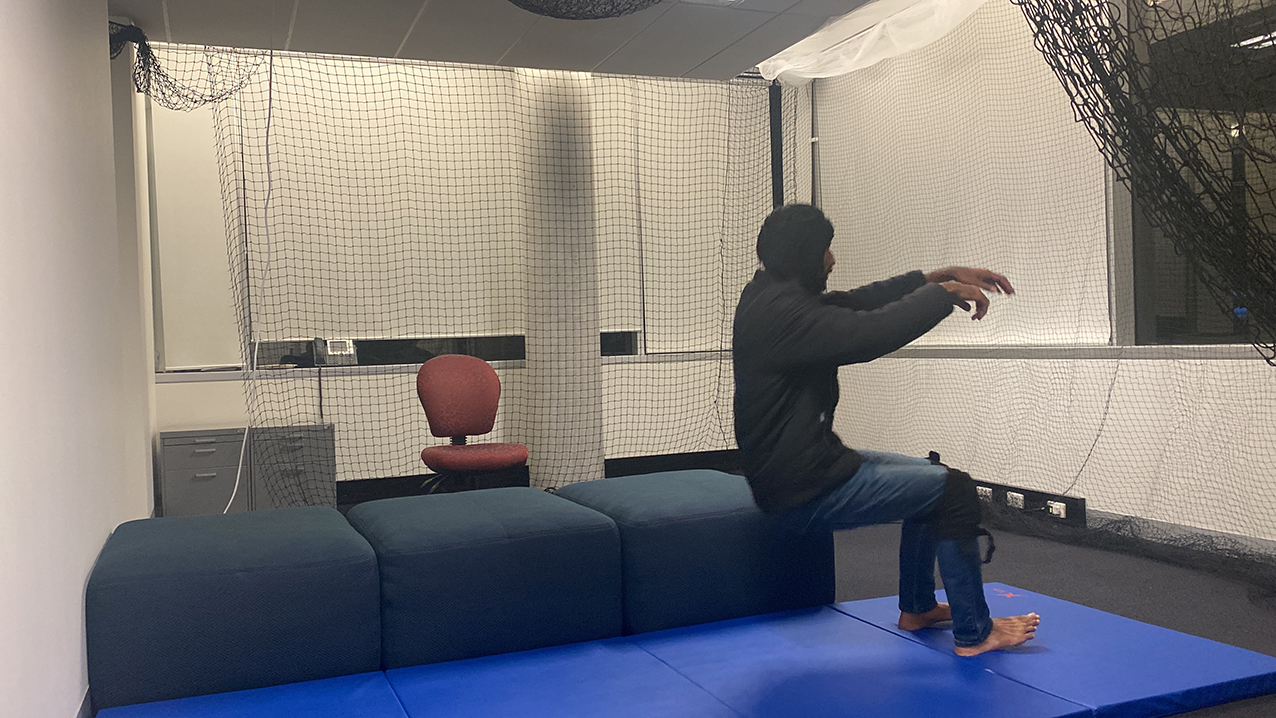}
        \caption{Fall: falling while sitting}
    \end{subfigure}
    \begin{subfigure}[b]{0.23\textwidth}
        \includegraphics[width=\textwidth]{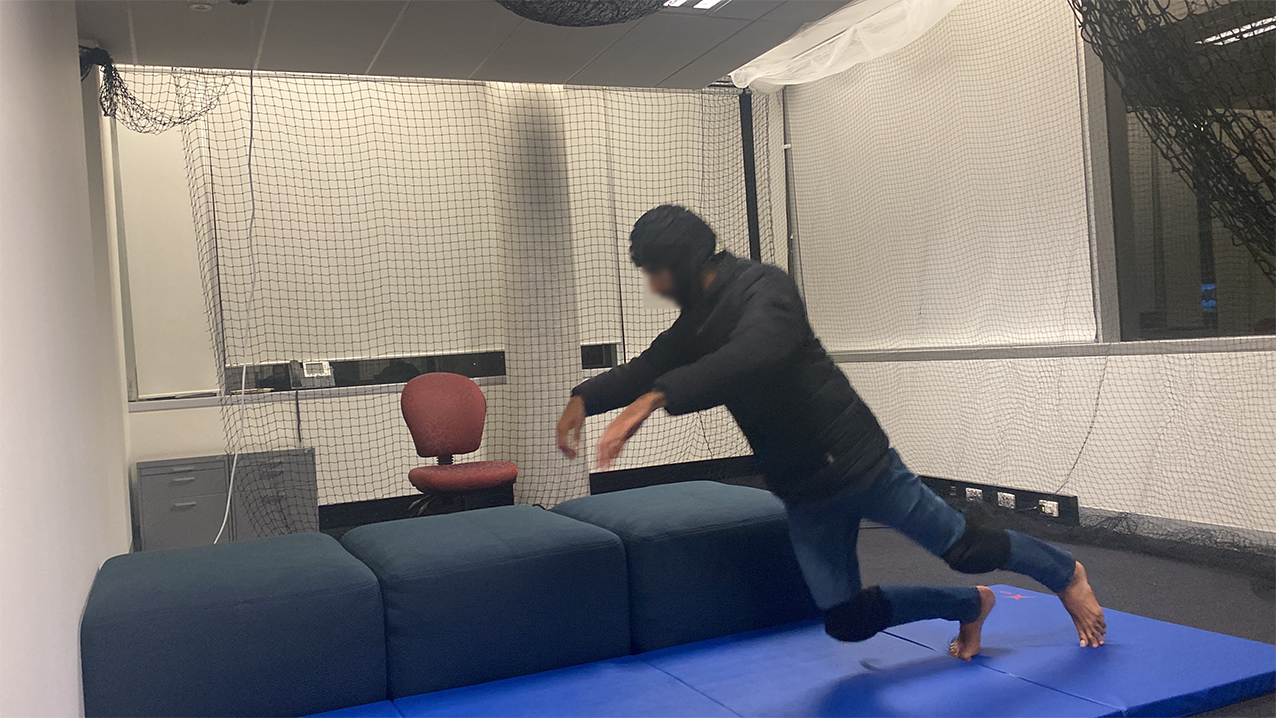}
        \caption{Fall: falling by tripping}
    \end{subfigure}
    \caption{Pictures of a subject performing experiments following the simulation protocol}
    \label{fig:photos_of_experiment}
\end{figure*}

\subsubsection{Simulation Protocol} \label{subsec:simulation_protocol}
A recent comprehensive fall dataset was published due to issues in existing publicly available wearable-based fall datasets \cite{FallAllD_dataset}.  
However, its protocol did not consider the behavioral characteristics of older adults or the common types of falls they experience in real-world settings. In our experiments, we focus on understanding the types of falls that elderly individuals typically experience in a smart home environment. For instance, overly strenuous ADLs such as jumping and running were excluded from our protocol, as most older adults do not engage in such high-intensity activities, especially at home. This approach ensures that the system is tailored to the expected characteristics of elderly users while effectively distinguishing between Activities of Daily Living (ADLs) and falls that may resemble certain ADLs. Table \ref{tab:simulation_protocol} summarizes our simulation protocol for falls and ADLs.

\subsubsection{Data Collection} \label{subsec:data_collection}

We conducted our experiments to collect data after setting up the experimental environment, configuring the UWB radar, and recruiting participants. Experiments were conducted with only the experimenter and subject in the room to ensure the noise of multiple people did not impact the data. Each subject cycled through the ADL and fall types shown in table \ref{tab:simulation_protocol}. Every volunteer repeated the experiment three times to minimize experimental error (see Fig.\ref{fig:process}). For each ADL/fall type, this was executed three times consecutively. A 10-second buffer between the start and end of each fall event adds up to a 20-second wait to allow the subject to reset their breathing and heart rate and limit overlaps between events. Some photos of a subject performing ADLs and falls are shown in Fig. \ref{fig:photos_of_experiment}.

\subsubsection{Data Preprocessing}
\begin{figure}[!t]
    \centering
    \includegraphics[width=0.6\linewidth]{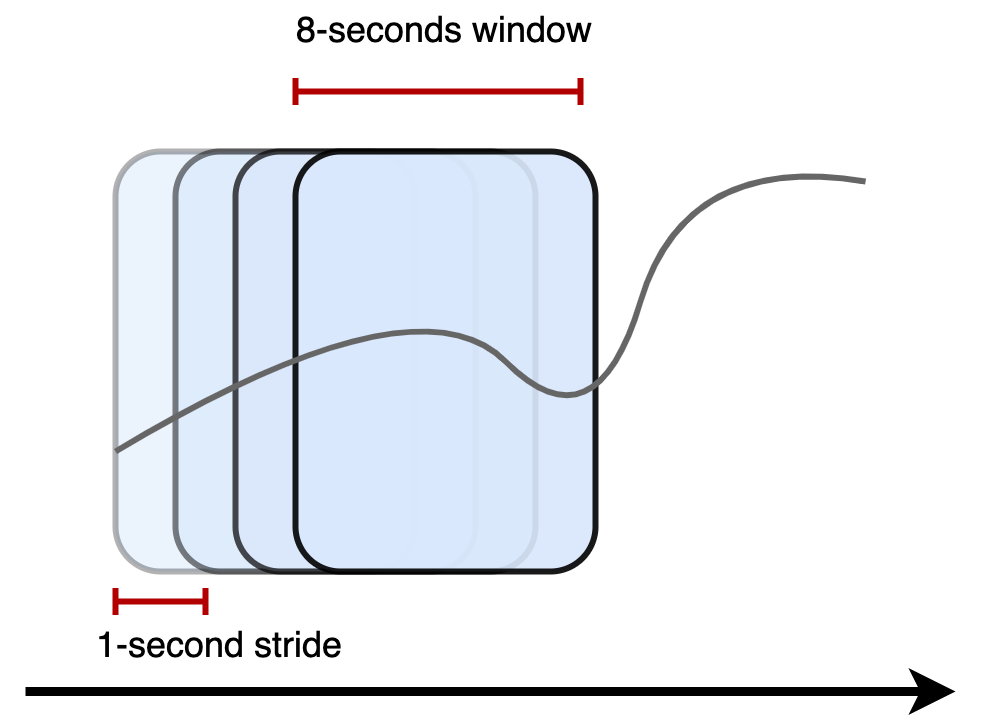}
    \caption{8-second sliding window with 1-second stride is used to extract the fall event time series. }
    \label{fig:sliding}
\end{figure}

The data collected to train the model was manually timestamped and labeled when an ADL or fall started, as well as the time it ended. After obtaining data,  the initial step is to exclude the invalid portion of the raw data (i.e., the data captured by the radar sensor during no activity). Then, we divide the valid data into independent samples. Note that different fall types have different actions leading up to the fall. These actions vary in length and need to be included in the classification. Even though deep learning models can handle variable-length inputs using masking or padding techniques, padding could potentially introduce noise into the training process \cite{lopez2020effect}. Therefore, we segment the data using a fixed window size to train the machine-learning models. We extracted an 8-second of data, i.e., four before and four after the event, into a sample (see Fig.\ref{fig:sliding}). In addition, we slid an 8-second window over the ADLs data with a stride of 1 second.

After sampling, we standardize the data as it comprises attributes with diverse value ranges, such as heartbeat, which slows the efficiency of the deep neural network in detecting patterns. During standardizing each attribute, the values were first subtracted by the mean and then divided by the standard deviation. To avoid the zero division problem, we added a small constant $\epsilon=1e^{-10}$ to the standard deviation term.

\textbf{Data Split} To train and test the performance of fall detection models, we randomly split the dataset into a training set (80\%), and a test set (20\%) while keeping the ratio of ADLs and fall samples unchanged. Then, we augmented the training set to generate new fall samples to balance the set as explained in Sec. \ref{data_aug}.

\begin{figure*}[!t]
\centering
\begin{subfigure}[b]{0.49\linewidth}
    \centering
    \includegraphics[width=\linewidth]{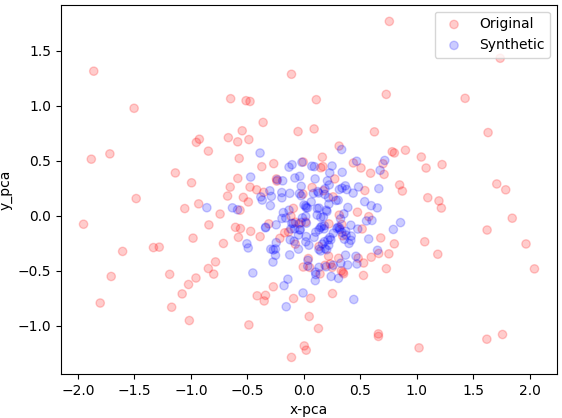}
    \caption{PCA plot}
    \label{fig:pca_plot}
\end{subfigure}
\hfill
\begin{subfigure}[b]{0.49\linewidth}
    \centering
    \includegraphics[width=\linewidth]{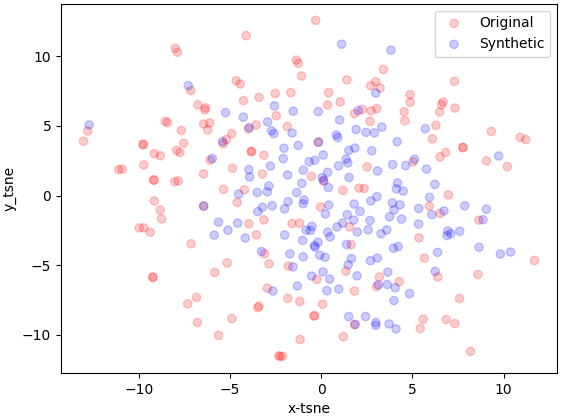}
    \caption{t-SNE plot}
    \label{fig:tsne_plot}
\end{subfigure}
\caption{The visualization for real and augmented fall data sequences generated by TTS-GAN}
\label{fig:synthetic_fall_samples_visualization}
\end{figure*}

\subsection{Hyperparameters Settings} 

\textbf{Hyperparameters for Data Augmentation:} As suggested in \cite{ENS_weighted_loss}, the hyperparameters $\beta$ and $\gamma$ of the weighted loss function were set to $0.9999$ and $2$, respectively. For the TTS-GAN model, we used the Adam optimizer with a batch size of 16 and a weight decay of $1e^{-3}$ \cite{Adam_paper}. The learning rate was set to $2e^{-4}$ for both the generator and discriminator. The model converged after 60 epochs.

\noindent \textbf{Hyperparameters for Fall Detection:} \space We also used Adam optimizer for fall detection models with a batch size of 64 and a weight decay of $1e^{-4}$ \cite{Adam_paper}. The learning rate was set to $1e^{-5}$ for all fall detection models. These classifiers were trained for 100 epochs.

\subsection{Metrics}

\subsubsection{Data augmentation Metrics}
To evaluate the dataset augmented by the GAN model, we can employ qualitative visualizations and quantitative metrics to check the similarity between real data and synthetic samples, as done in \cite{TTS_GAN}. To illustrate this similarity, we plotted example graphs of data distributions transformed to two dimensions using PCA and t-SNE. In quantitative evaluation, we selected average cosine similarity as the numerical metric, as defined in the appendix of \cite{TTS_GAN}. Before calculating the average cosine similarity, we need to extract seven statistical features from every signal sequence for each attribute: mean, median, standard deviation, variance, root mean square, minimum, and maximum.

Suppose we extract $m$ features from all attributes of each sample and form a feature vector in the format $f = <feature_1, feature_2, feature_3,..., feature_m>$. Then, for every pair of real data sequence feature vector $f_a$ and artificial data sequence feature vector $f_b$, the cosine similarity is defined as below:

\begin{equation}
cos_sim_{ab} = \frac{f_a \cdot f_b}{||f_a||||f_b||} = \frac{\Sigma_{i=1}^m f_{ai}f_{bi}}{ \sqrt{\Sigma_{i=1}^m f_{ai}^2} \sqrt{\Sigma_{i=1}^m f_{bi}^2}}
\end{equation}

The average cosine similarity score is the arithmetic mean of cosine similarity between pairs of vector features belonging to real and synthetic data signals of the same class. This can be calculated as follows, where $n$ is the total number of samples:

\begin{equation}
avg_cos_sim = \frac{1}{n}\Sigma_{i=1}^n cos_sim_i
\label{eq_cos}
\end{equation}

\subsubsection{Fall Detection Metrics} Most fall detection systems in the literature use accuracy as the evaluation metric. However, since our dataset is extremely unbalanced (i.e., most samples are ADLs), metrics for imbalanced binary classification, such as Balanced Accuracy and F1-score, are preferred to evaluate classifier performance. To compute Balanced Accuracy and F1-score, we must first calculate sensitivity and specificity. Sensitivity measures how many falls are correctly detected, and specificity considers how many ADLs are correctly classified. These can be defined as follows:

\begin{equation}
sensitivity = \frac{TP}{TP+FN}
\end{equation}

\begin{equation}
specificity = \frac{TN}{TN+FP}
\end{equation}

\noindent Using sensitivity and specificity, we then calculate the F1-score and Balanced Accuracy:

\begin{equation}
F1-score = \frac{2 \cdot sensitivity \cdot specificity}{sensitivity + specificity}
\end{equation}

\begin{equation}
Balanced\_Accuracy = \frac{sensitivity + specificity}{2}
\end{equation}

\subsection{Evaluation}


In the first experiment, we trained the TTS-GAN model using the training set to generate fall data. Fig.  \ref{fig:synthetic_fall_samples_visualization} shows the distribution of real (red) and synthetic (blue) data points mapped to two-dimensional space using PCA and t-SNE. The distribution patterns of actual and synthetic data are similar, with significant overlap in the PCA plot. The average cosine similarity score is computed using Eq.\ref{eq_cos}. The average cosine similarity score between real and synthetic samples is 0.8513, indicating a high similarity.



\begin{figure}[!t]
    \centering
    \includegraphics[width=\linewidth]{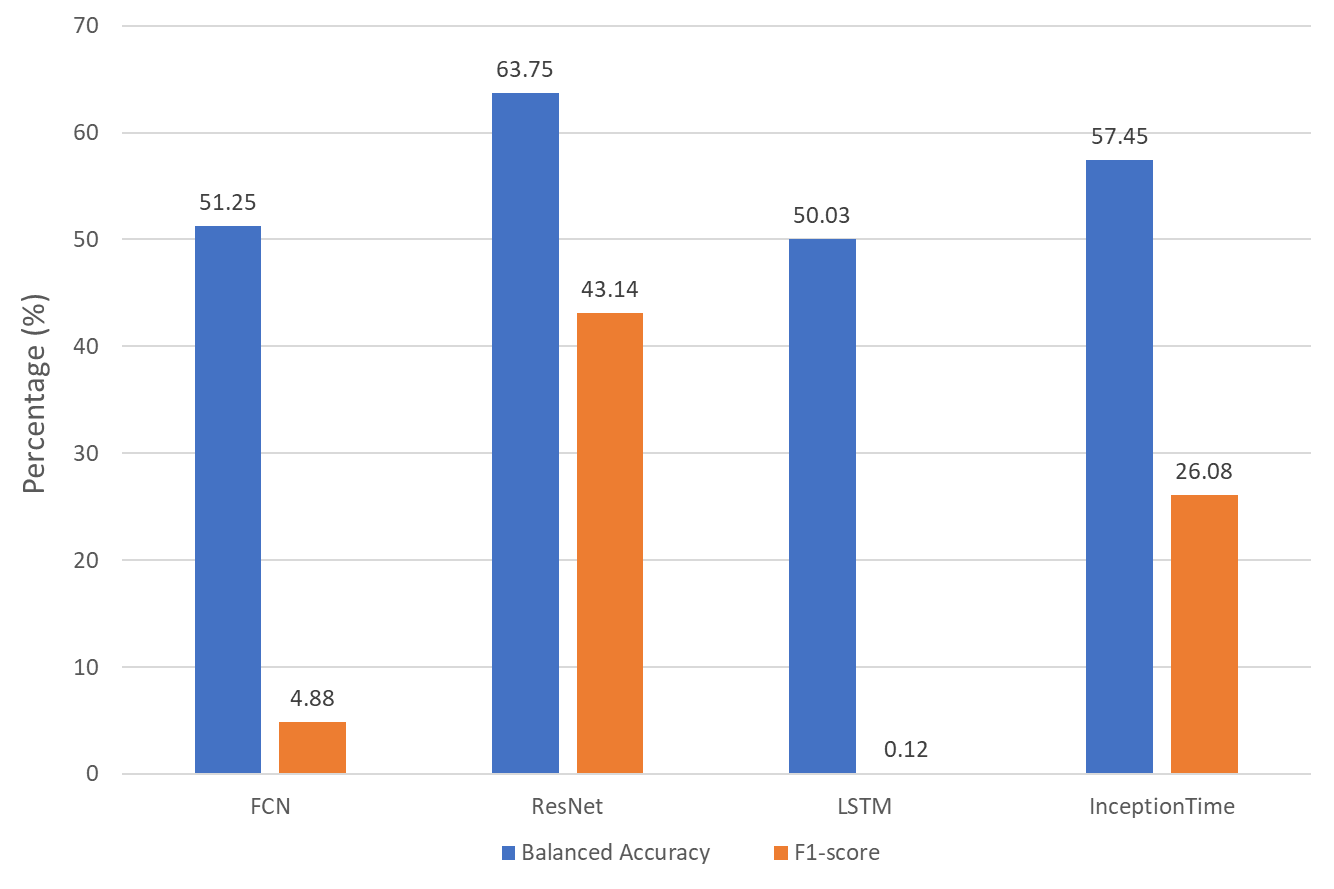}
    \caption{The performance of fall detection services without data augmentation}
    \label{fig:none_performance}
\end{figure}

\begin{figure}[!t]
    \centering
    \includegraphics[width=\linewidth]{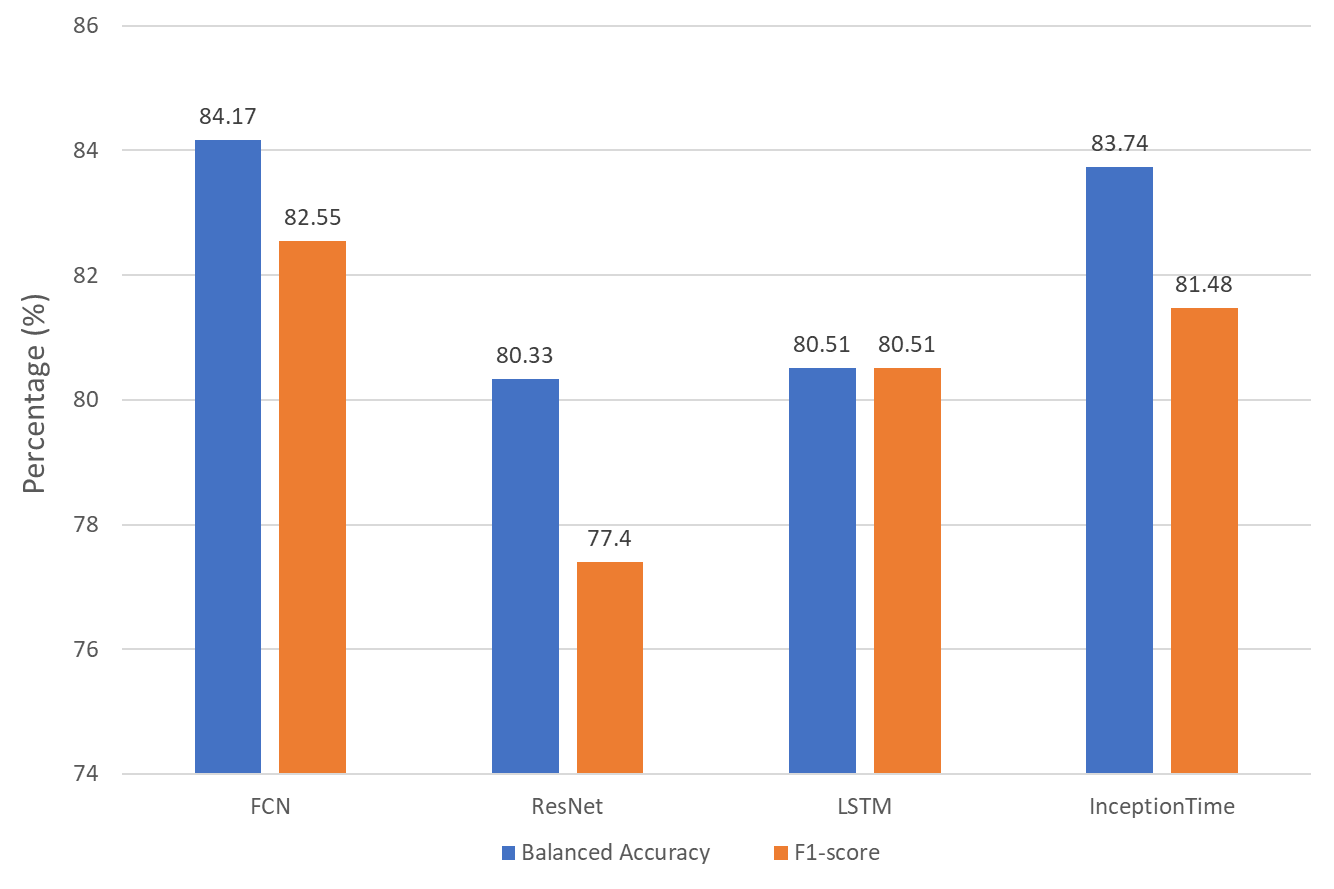}
    \caption{The performance of fall fall detection services with using SMOTE for data augmentation}
    \label{fig:SMOTE_performance}
\end{figure}

\begin{figure}[!t]
    \centering
    \includegraphics[width=\linewidth]{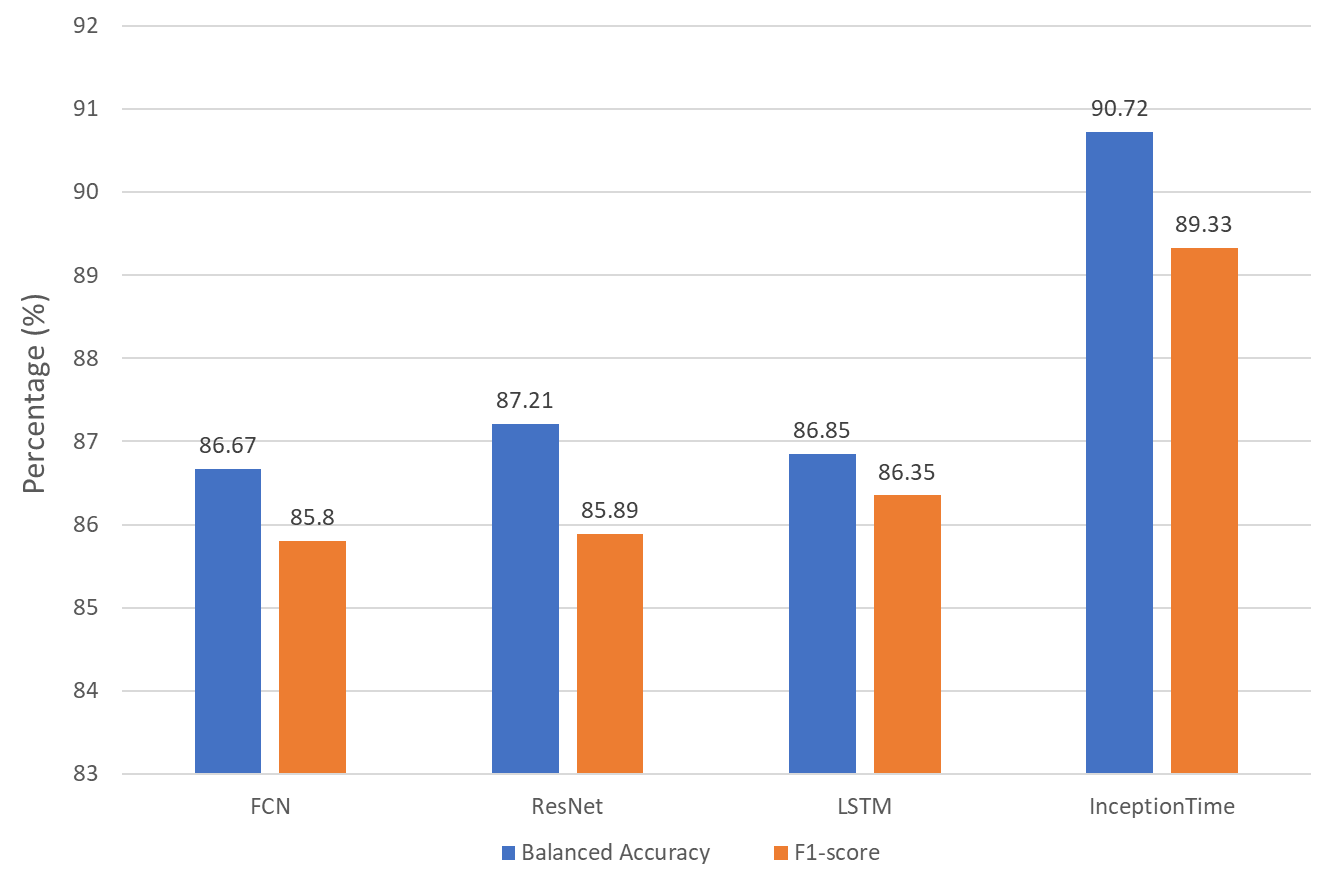}
    \caption{The performance of fall detection models using TTS-GAN for data augmentation}
    \label{fig:TTS_GAN_performance}
\end{figure}

 The performance of all fall detection models using different techniques for dealing with data imbalance is summarized in Tables \ref{tab:overall_balanced_accuracy_table} and Table \ref{tab:overall_F1_score_table}. Additionally, the performance of all models without any data imbalance strategy, with the SMOTE technique, and employing the TTS-GAN model is visualized in Fig. \ref{fig:none_performance}, Fig. \ref{fig:SMOTE_performance}, and Fig. \ref{fig:TTS_GAN_performance}, respectively. We can observe two phenomena from these summary tables and figures.

Tables in Tables \ref{tab:overall_balanced_accuracy_table} and  \ref{tab:overall_F1_score_table} summarize the performance of all fall detection models using different techniques to address data imbalance. Additionally, Figs \ref{fig:none_performance}, \ref{fig:SMOTE_performance}, and \ref{fig:TTS_GAN_performance} visualize the performance of models as: (1) without any data imbalance strategy, (2) with the SMOTE technique, and (3) employing the TTS-GAN model, respectively. Two phenomena can be observed from these summary tables and figures.

First, the TTS-GAN model significantly improves the performance of fall detection models compared to other augmentation methods. This is because the TTS-GAN model approximates the distribution of fall samples and resamples synthetic fall sequences, adding new knowledge to the training set. This enables the classifier to explore hidden patterns more effectively. While the SMOTE technique also generates artificial fall samples, these samples are created by slightly moving existing fall samples in high-dimensional space. If some test set fall samples are not close to the training set samples, SMOTE’s effectiveness is limited. This results in minimal performance improvement. In contrast, the TTS-GAN model can generate samples closer to test set samples. This greatly improves model performance as long as it does not overfit on the training set.

\begin{table}[!t]
\centering
\caption{Balanced Accuracy of all models and data augmentation methods}
\resizebox{\linewidth}{!}{%
\begin{tabular}{@{}ccccccc@{}}
\toprule
\textbf{} &
  \textbf{None} &
  \textbf{INS} &
  \textbf{ENS} &
  \textbf{ROS} &
  \textbf{SMOTE} &
  \textbf{TTS-GAN} \\ \midrule
\textbf{FCN}           & 51.25                     & 78.36 & 81.08 & 80.90 & 84.17 & 86.67          \\
\textbf{ResNet}        & 63.75                     & 84.69 & 81.97 & 80.78 & 80.33 & 87.21          \\
\textbf{LSTM}          & \multicolumn{1}{l}{50.03} & 70.19 & 72.91 & 84.86 & 80.51 & 86.85          \\
\textbf{InceptionTime} & \multicolumn{1}{l}{57.45} & 88.24 & 87.05 & 83.86 & 83.74 & \textbf{90.72} \\ \bottomrule
\end{tabular}%
}
\label{tab:overall_balanced_accuracy_table}
\end{table}

\begin{table}[!t]
\centering
\caption{F1 score evaluation of all models and data augmentation methods }
\resizebox{\linewidth}{!}{%
\begin{tabular}{@{}ccccccc@{}}
\toprule
\textbf{} &
  \textbf{None} &
  \textbf{INS} &
  \textbf{ENS} &
  \textbf{ROS} &
  \textbf{SMOTE} &
  \textbf{TTS-GAN} \\ \midrule
\textbf{FCN}           & 4.88                      & 78.35 & 80.62 & 80.03 & 82.55 & 85.80          \\
\textbf{ResNet}        & 43.14                     & 83.58 & 80.88 & 79.34 & 77.40 & 85.89          \\
\textbf{LSTM}          & \multicolumn{1}{l}{0.12}  & 70.19 & 70.63 & 84.78 & 80.51 & 86.35          \\
\textbf{InceptionTime} & \multicolumn{1}{l}{26.08} & 87.72 & 86.70 & 82.32 & 81.48 & \textbf{89.33} \\ \bottomrule
\end{tabular}%
}
\label{tab:overall_F1_score_table}
\end{table}

The second observation is that the InceptionTime model generally outperforms the other three baseline models in fall detection. This is likely because the parallel convolutional layers with different kernel sizes in the InceptionModule enable the InceptionTime model to automatically and effectively extract discriminative patterns at different time scales from the data.\looseness=-1

\section{Related works}
The study of fall detection systems lies at the intersection of technology and deep learning. As our society ages, the need for systems that can autonomously monitor and respond to the health needs of the elderly, particularly in detecting falls, has grown. This section discusses recent advancements in fall detection systems technologies \cite{Ramachandran2020}. Additionally, we explore the latest deep-learning solutions for fall detection.
\subsection{Fall Detection Systems Technologies}

This section discusses recent advancements in fall detection systems technologies, categorized by the sensor types used: wearable, vision-based, and ambiance-based detection methods.\looseness=-1

\subsubsection{Wearable-based detection systems} Wearable-based systems utilize a combination of sensors such as accelerometers, gyroscopes, and magnetometers attached to the patient's body. These devices monitor vital signs and movements, alerting caregivers through connected mobile devices upon detecting a fall. Despite their portability and low power consumption, their effectiveness hinges on continuous wear and regular charging, posing a challenge for consistent usage \cite{ozdemir2016analysis} \cite{Ramachandran2020} \cite{yusoff2022towards}. Recent shifts towards using built-in sensors in smartphones and smartwatches aim to address these challenges, although they are not without limitations related to device battery life and the habit of carrying these devices consistently.


\subsubsection{Vision-based detection systems} Vision-based systems leverage image processing techniques and deep learning models to detect falls from different camera types, including RGB and depth cameras. These systems benefit from multiple viewpoints to reduce false alarms, applying advanced neural networks for high precision and recall rates \cite{shojaei2018video} \cite{nunez2017vision} \cite{ozcan2016autonomous}. However, the broad adoption of such systems is curtailed by significant privacy concerns and the challenge of data accessibility for patients.


\subsubsection{Radar-based detection systems}


Radar-based sensors, including Continuous Wave (CW), Frequency-Modulated Continuous-Wave (FMCW), and Ultra-Wideband (UWB) radars, have gained traction in fall detection due to their ability to monitor movement and vital signs without requiring physical contact \cite{erol2017range} \cite{shrestha2019human} \cite{sadreazami2020fall} \cite{maitre2020fall}. UWB radars, in particular, provide high spatial resolution and privacy-preserving monitoring by emitting short pulses at high frequencies \cite{wang2019uwb}. However, despite their advantages, existing radar-based fall detection methods face significant challenges. Many studies rely on limited datasets, often involving a small number of participants in controlled environments, which limits their generalizability \cite{maitre2020fall}. Another study developed the WITSCare smart home system to help elderly individuals live independently \cite{yao2016context}. The system detects IoT-enabled devices related to various contexts such as the residents' locations, activities, and interactions with home appliances, and abstracts these devices as services. However, the study did not conduct experiments on fall detection.  Our research extends this domain by exploring data augmentation through TTS-GAN models to enhance detection accuracy in diverse settings.



\subsection{Deep learning for Fall Detection Systems }

The application of deep learning in fall detection has opened new avenues for accurate and efficient systems \cite{sadreazami2019fall} \cite{wang2017time}. CNNs and ResNets have demonstrated remarkable success in classifying time-series data from sensors, capturing the intricate patterns of falls \cite{sadreazami2019fall} \cite{wang2017time}. Similarly, LSTM models have enhanced the processing of sequential data, providing a nuanced understanding of the temporal dynamics critical for fall detection \cite{maitre2020fall} \cite{Uddin2020} \cite{ozdemir2016analysis}. InceptionTime models, known for their innovative architecture, offer a promising direction for time-series classification in fall detection. These models have achieved high accuracy rates by extracting latent hierarchical features \cite{wang2022near}.


Machine learning and deep learning models often have a tendency to classify samples as the majority class\cite{fawaz2018data}. This may reduce the performance of fall detection frameworks. Data augmentation techniques have been used to address imbalanced time series classification \cite{fawaz2018data}. However,  most successful data augmentation applications have been in image classification, which often overlooks the temporal dependencies between features. Although recent studies have explored data augmentation for time series classification and imbalanced classes, the improvements have been minimal \cite{iwana2021empirical}. Further research is necessary to fully understand how these techniques impact model performance.




\section{Conclusion}

In this study, we introduced a novel IoT-based service-oriented architecture for fall detection in smart home environments aimed at promoting independent living for the elderly. By integrating UWB radar sensors with the Fall Detection Generative Pre-trained Transformer (FD-GPT), our system effectively addresses the challenges of data scarcity and accuracy. The results from our experiments highlight the framework's ability to accurately detect falls and distinguish them from ADLs, ensuring timely interventions and enhancing the safety and quality of life for elderly residents. This research underscores the potential of unobtrusive sensing technologies in developing practical and reliable smart aging solutions.

\section*{Acknowledgment}
This research was partly made possible by LE220100078 and DP220101823 grants from the Australian Research Council. The statements made herein are solely the responsibility of the authors.

\bibliographystyle{IEEEtran}
\bibliography{main}

\end{document}